\documentclass[twocolumn,showpacs,amsmath,amssymb,superscriptaddress,prl]{revtex4}
\usepackage{graphicx}
\usepackage{dcolumn}
\usepackage{bm}
\usepackage{indentfirst}
\begin{document}
\title{Singlet-Triplet Physics and Shell Filling in Carbon Nanotube Double Quantum Dots}
\author{H. Ingerslev J\o rgensen}
\email{hij@fys.ku.dk} \affiliation{Hitachi Cambridge laboratory,
Hitachi Europe Ltd., Cambridge CB3 0HE, United Kingdom}
\affiliation{Nano-Science Center, Niels Bohr Institute, University
of Copenhagen, Universitetsparken 5, DK-2100~Copenhagen \O ,
Denmark}
\author{K. Grove-Rasmussen}
\affiliation{Nano-Science Center, Niels Bohr Institute, University
of Copenhagen, Universitetsparken 5, DK-2100~Copenhagen \O ,
Denmark}
\author{K.-Y. Wang}
\affiliation{Hitachi Cambridge laboratory, Hitachi Europe Ltd.,
Cambridge CB3 0HE, United Kingdom}
\author{A.~M.~Blackburn}
\affiliation{Hitachi Cambridge laboratory, Hitachi Europe Ltd.,
Cambridge CB3 0HE, United Kingdom}
\author{K. Flensberg}
\affiliation{Nano-Science Center, Niels Bohr Institute, University
of Copenhagen, Universitetsparken 5, DK-2100~Copenhagen \O ,
Denmark}
\author{P. E. Lindelof}
\affiliation{Nano-Science Center, Niels Bohr Institute, University
of Copenhagen, Universitetsparken 5, DK-2100~Copenhagen \O ,
Denmark}
\author{D. A. Williams}
\affiliation{Hitachi Cambridge laboratory, Hitachi Europe Ltd.,
Cambridge CB3 0HE, United Kingdom}
\date{\today}
%
\maketitle
\textbf{
An artificial two-atomic molecule, also called a double quantum dot
(DQD), is an ideal system for exploring few electron physics
\cite{petta01,koppens,johnson_nature,johnson,wiel,pfund,ono,elzerman,fuhrer,fasth,chan,hayashi,hansonreview,nowack}.
Spin-entanglement between just two electrons can be explored in such
systems where singlet and triplet states are accessible.
These two spin-states can be regarded as the two states in a quantum
two-state system, a so-called singlet-triplet qubit
\cite{nielsenchuang}.
A very attractive material for realizing spin based qubits is the
carbon nanotube (CNT)
\cite{mason,biercuk2,sapmaz,graber,graber2,jorgensen}, because it is
expected to have a very long spin coherence time
\cite{gorman,cain,hu,coish,chan}.
Here we show the existence of a gate-tunable singlet-triplet qubit
in a CNT DQD. We show that the CNT DQD has clear shell structures of
both four and eight electrons, with the singlet-triplet qubit
present in the four-electron shells.
We furthermore observe inelastic cotunneling via the singlet and
triplet states, which we use to probe the splitting between singlet
and triplet, in good agreement with theory.
}
\newline \indent
Creating a qubit in a solid-state system demands control of the
number of interacting electrons. This control has to date been
obtained using semiconducting materials operated close to the
band-gap edge. We show in this Letter that shell structures in CNT
DQDs, owing to the 1-dimensional nature of the CNT, can be used to
obtain the same kind of control. Both 4-electron and 8-electron
(DQD)-shells are observed.
We use one of the 4-electron shells to entangle the spin of two
electrons and show that by separating the two electrons into
separate QDs they form a spin triplet state, and by collecting them
into the same QD they form a spin singlet state, i.e., a
gate-tunable singlet-triplet qubit.
\newline \indent
\begin{figure}[b!]
\begin{center}
\includegraphics[width=0.45\textwidth]{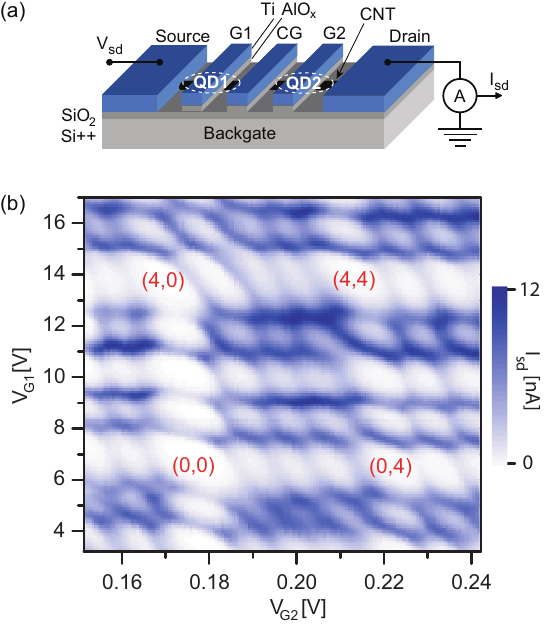}
\end{center}
\caption{\textbf{Shell filling of a carbon nanotube double quantum
dot.} (a) Schematic figure of the device consisting of a carbon
nanotube (CNT) contacted by titanium source and drain electrodes and
gated with three top-gate electrodes, G1, CG (center gate), and G2.
Two quantum dots (QD1, QD2) are formed in series, one under G1 and
one under G2. (b) Surface plot of current ($I_{\rm sd}$) at constant
bias ($V_{\rm sd}=0.5$\,mV) as function of voltage applied to G1
($V_{\rm G1}$) and G2 ($V_{\rm G2}$). Red numbers (N,M) are shell
occupation numbers for one 8-electron (DQD)-shell. Two further
8-electron shells were observed in connection the this shell, one
below, and one to the right. \label{fig:fig1}}
\end{figure}
The device analyzed in this Letter, schematically shown in
Fig.~\ref{fig:fig1}(a), is comprised of a CNT contacted by titanium
electrodes, and gated by three top-gate electrodes, G1, CG
(center-gate), and G2, made of aluminum oxide and titanium. The
device has two strongly coupled quantum dots in series as confirmed
by the observation of the so-called honeycomb pattern in current
($I_{\rm sd}$) versus voltage applied to G1 ($V_{\rm G1}$) and G2
($V_{\rm G2}$) (Fig.~\ref{fig:fig1}(b) and Fig.~\ref{fig:fig2}(a))
\cite{wiel}.
The tunneling barrier between the two dots is due to a defect in the
CNT (similar to Ref.~[\onlinecite{mason}]).
The resulting two dots have roughly equal charging energies and
level spacings (see below), from which we infer that the defect is
located under or close to the center gate.
%
%
The number of electrons in dot 1 and dot 2 can be controlled by
tuning $V_{\rm G1}$ and $V_{\rm G2}$, respectively. In the middle of
each hexagon (white areas in Fig.~\ref{fig:fig1}(b) and
Fig.~\ref{fig:fig2}(a)) a fixed number of electrons are localized in
each dot, and electron transport is suppressed by Coulomb blockade.
Along the entire edge of the hexagons (blue lines), single electron
transport is allowed through molecular states formed in the DQD,
indicating a strong coupling between the two dots. The height
(width) of the hexagons corresponds to the energy required to add an
extra electron in dot 1 (dot 2). In Fig.\ref{fig:fig1}(b) the width
and height of the hexagons alternate in size in a regular pattern.
The four hexagons marked with red numbers are distinctively larger
than the other hexagons with three smaller hexagons in between,
indicating that each dot has four-fold degenerate levels due to spin
and orbital degeneracy \cite{nygaardQD,moriyama_prl,buitelaar02}. An
8-electron shell structure of the DQD can therefore be identified in
this plot. Shell occupation numbers (N,M), where N (M) is the level
occupation number in dot 1 (dot 2) are written onto the honeycomb
diagram.
\begin{figure}
\begin{center}
\includegraphics[width=0.45\textwidth]{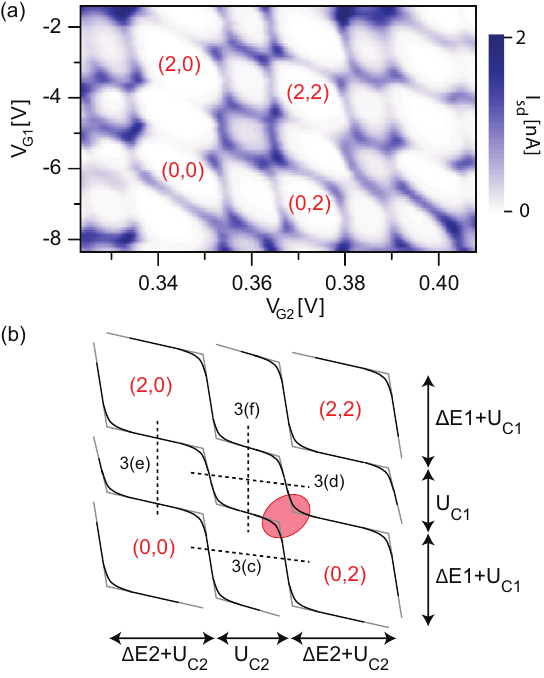}
\end{center}
\caption{\textbf{Four-electron DQD-shell structure.} (a) Surface
plot of current ($I_{\rm sd}$) at constant bias ($V_{\rm
sd}=0.2$\,mV) as function of voltage applied to G1 ($V_{\rm G1}$)
and G2 ($V_{\rm G2}$). The numbers (N,M) are shell occupation
numbers for one 4-electron shell. (b) Black lines: Schematic
honeycomb diagram for a 4-electron shell with strong tunnel
coupling, and a small cross capacitance. Gray lines: Same honeycomb
diagram with negligible tunnel coupling. Dashed lines indicate where
the line-traces in Fig.~\ref{fig:fig3} are measured, and the red
oval indicates the region analyzed in Fig.~\ref{fig:fig4}.
\label{fig:fig2}}
\end{figure}
\begin{figure}
\begin{center}
\includegraphics[width=0.48\textwidth]{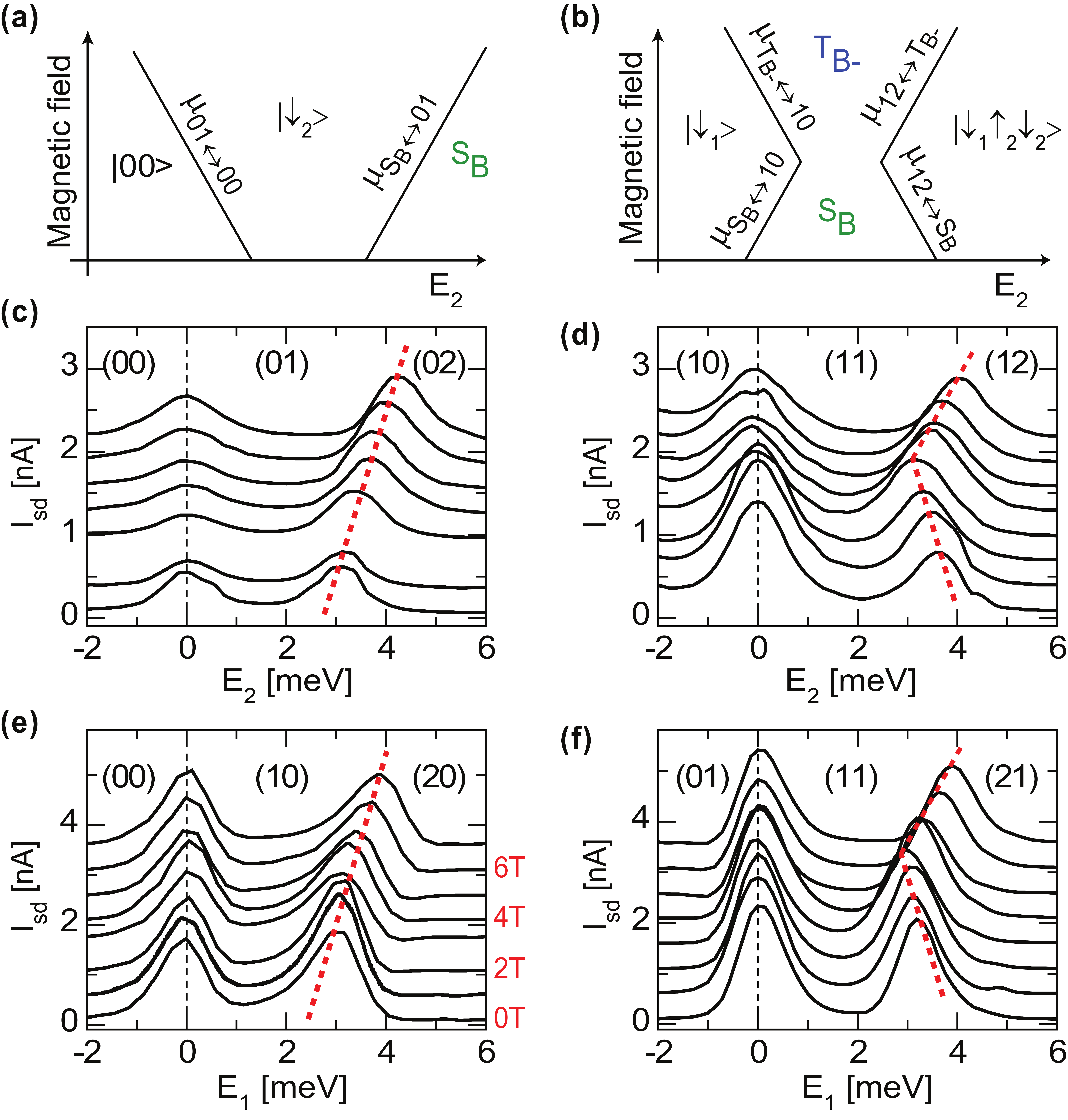}
\end{center}
\caption{\textbf{Singlet and Triplet states in a four-electron
shell.}
(a)-(b) Theoretical magnetic field dependence of the chemical
potentials measured in (c)-(f). In (a) the 4-electron shell is
filled from zero electrons ($|00\rangle$), to one spin-down electron
in dot 2 ($\mid \downarrow_2 \rangle$), to a singlet (${\rm S_B}$).
In (b) the device is filled from a spin-down electron in dot 1
($\mid \downarrow_1 \rangle$), to either a singlet (${\rm S_B}$) or
a triplet (${\rm T_{B-}}$) (see text), to a three particle state
with one spin-down electron in dot 1 and both a spin-down and
spin-up electron in dot 2 ($\mid \downarrow_1 \uparrow_2
\downarrow_2 \rangle$).
The line traces in (c)-(f) are extracted from honeycomb diagrams
measured at $B=0,1,2,...,7$\,T, where $B$ is perpendicular to the
nanotube.
(c) and (e) Horizontal and vertical line traces through hexagon
(0,1) and (1,0) as function of electrostatic potential in dot 2
($E_2$) and dot 1 ($E_1$), respectively.
(d) and (f) Horizontal and vertical line traces through hexagon
(1,1).
Each line in (c)-(f) is offset 0.2\,nA, 0.3\,nA, 0.3\,nA, and
0.5\,nA respectively for clarity, and the left-most peak is centered
at zero.
\label{fig:fig3}}
\end{figure}
\newline \indent
The honeycomb diagram in Fig.~\ref{fig:fig2}(a) is measured for the
same device but in another gate region where a new pattern in the
sizes of the hexagons is observed. The hexagons alternate in size
between large and small due to only spin degeneracy of the energy
levels in each dot \cite{nygaardQD,moriyama_apl}, yielding a
4-electron shell structure of the DQD. The charging energies
($U_{\rm C1}$, $U_{\rm C2}$) and level spacings ($\Delta E_1$,
$\Delta E_2$) for the two dots can be extracted from the honeycomb
pattern as schematically shown in Fig.~\ref{fig:fig2}(b). The gate
coupling of G1 (G2) to dot 1 (dot 2) is found from bias spectroscopy
plots (not shown), and we find $U_{\rm C1} \sim 3$\,meV and $\Delta
E_1 \sim 1.2$\,meV for dot 1, and $U_{\rm C2} \sim 3.5$\,meV and
$\Delta E_2 \sim 1.5$\,meV for dot 2. Since charging energy and
level spacing are almost identical for the two dots we deduce that
the two dots are roughly equal in size. We have observed both
4-electron and 8-electron shell structures in two different devices.
\newline \indent
\begin{figure*}
\begin{center}
\includegraphics[width=0.7\textwidth]{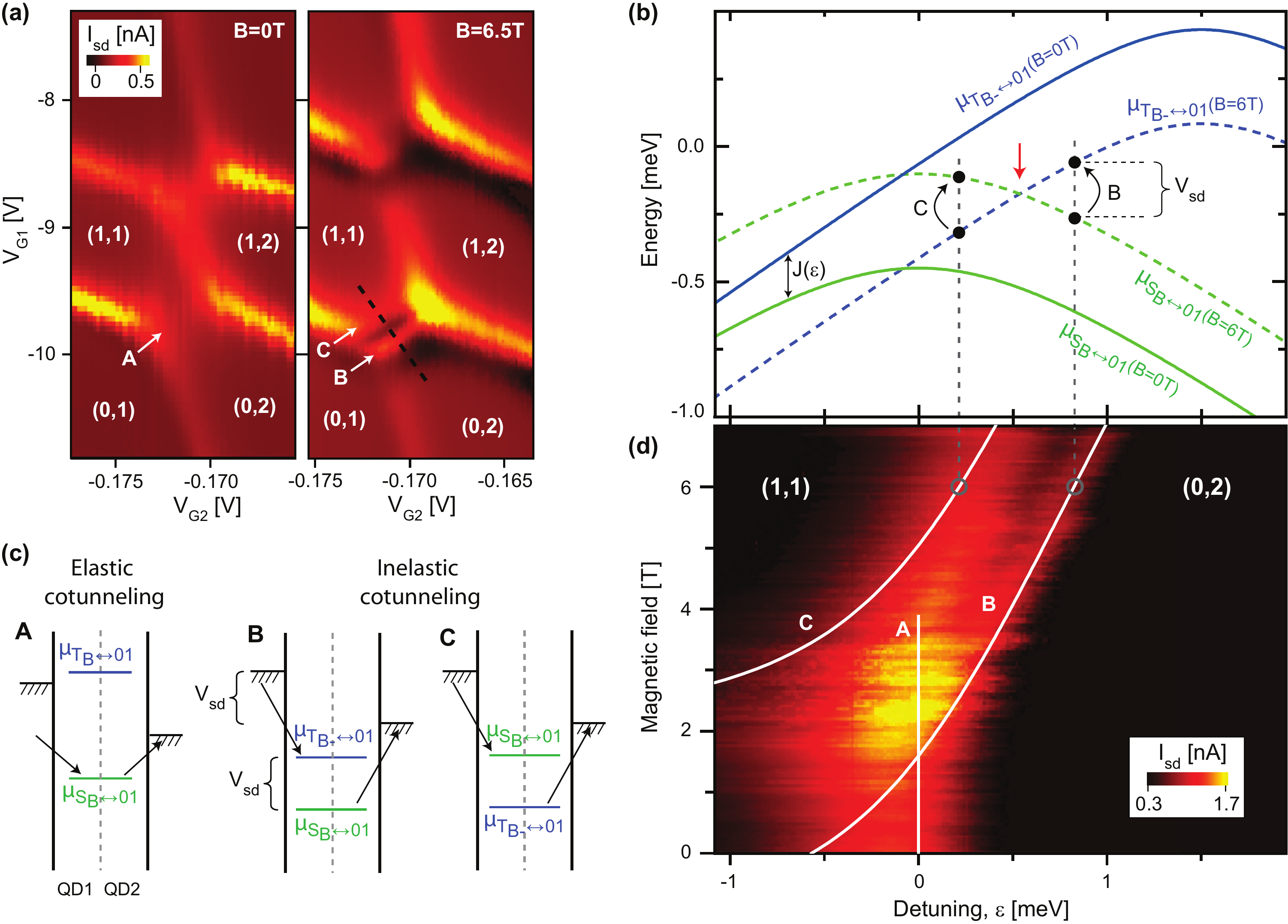}
\end{center}
\caption{\textbf{Singlet-triplet splitting probed by inelastic
cotunneling.}
(a) Small section of the honeycomb diagram analyzed in
Fig.~\ref{fig:fig3} with $V_{\rm sd} = 50 \mu$V at $B=0$\,T (left)
and $B=6.5$\,T (right). The numbers (N,M) indicate electron
occupation of the 4-electron shell.
(b) Chemical potentials for the singlet bonding ($\mu_{{\rm
S_B}\leftrightarrow01}$) and triplet bonding ($\mu_{{\rm
T_{B-}}\leftrightarrow01}$) with $B=0$\,T (solid green and blue
lines) and with $B=6$\,T (dashed green and blue lines) calculated
using Eq.~(\ref{eq:mu_s_b}) and (\ref{eq:mu_t_b}).
Onset of inelastic cotunneling which excites electrons from singlet
to triplet (black arrow marked B) and from triplet to singlet (black
arrow marked C) occurs when the separation between $\mu_{{\rm
S_B}\leftrightarrow01}$ and $\mu_{{\rm T_{B-}}\leftrightarrow01}$ is
equal to $e V_{\rm sd}$.
(c) Schematic transport diagrams for elastic cotunneling (A) and
inelastic cotunneling (B and C).
(d) Surface plot of current ($I_{\rm sd}$) at $V_{\rm sd} = 0.2$\,mV
versus magnetic field ($B$), and detuning ($\varepsilon$) along the
black dashed line in (a).
Onset of inelastic cotunneling occurs along the white lines marked B
and C, calculated using Eq.~(\ref{eq:condition}) with $t=0.32$\,meV,
$\Delta E_2=1.5$\,meV, $V_{\rm sd}=0.2$\,mV, and $g=2$.
Dashed grey lines indicate where the two inelastic cotunneling
processes shown in (b) occurs.
 \label{fig:fig4}}
\end{figure*}
We will in the rest of the Letter focus on a 4-electron shell with
level spacings and charging energies similar the the 4-electron
shell shown in Fig.~\ref{fig:fig2}(a), except $\Delta E_1 \sim
1.9$\,meV.
The singlet ground state between region (1,1) and (0,2) is in
general a bonding state of the local singlet (S(02), both electrons
in dot 2), and the nonlocal singlet (S(11), one electron in each
dot):
\begin{equation}
{\rm S_{B}} = \alpha {\rm S(11)} + \beta {\rm S(02)}
\end{equation}
The detuning ($\varepsilon = E_2 - E_1$) dependent parameters
$\alpha$ and $\beta$ determine the weight of each state, and $E_1$
and $E_2$ are the electrostatic potentials in dot 1 and dot 2,
respectively. Similarly for the triplets
\begin{equation}
\begin{array}{lcl}
{\rm T_{B-}} & = & \alpha' {\rm T_-(11)} + \beta' T_-(02) \\
{\rm T_{B0}} & = & \alpha' {\rm T_0(11)} + \beta' T_0(02) \\
{\rm T_{B+}} & = & \alpha' {\rm T_+(11)} + \beta' T_+(02)
\end{array}
\end{equation}
where $-,0,+$ denotes the spin magnetic moment in the z-direction,
$S_{\rm z}=-1,0,+1$.
We will in the following show the existence of the singlet and
triplet states, i.e., the singlet-triplet qubit, on the basis of a
magnetic field spectroscopy on the 4-electron shell.
\newline \indent
In Fig.~\ref{fig:fig3}(c) we analyze the magnetic field dependence
of the width of hexagon (0,1), which involves 0, 1, and 2 electrons
in dot 2, The chemical potential for these two Coulomb peaks is
given by \cite{hansonreview}: $\mu_{01\leftrightarrow 00} \propto
-\frac{1}{2} g\mu_{\rm B} B$ and $\mu_{\rm S_B\leftrightarrow 01}
\propto +\frac{1}{2} g\mu_{\rm B} B$, where $\mu_{01\leftrightarrow
00}$ is the chemical potential for adding an electron to charge
state (01) given no electron in the DQD-shell, and $\mu_{\rm
S_B\leftrightarrow 01}$ is the chemical potential for adding an
electron in state ${\rm S_B}$ given one electron charge state (01).
These two Coulomb peaks are therefore expected to separate by $g
\mu_{\rm B} B$ as shown in Fig.~\ref{fig:fig3}(a). The height of
hexagon (1,0) are analogously expected to separate by $g \mu_{\rm B}
B$. The measurements in Fig.~\ref{fig:fig3}(c) and (e) are in good
quantitative agreement with the theory in Fig.~\ref{fig:fig3}(a).
The measured separation at 7\,T is 0.95\,meV and 0.8\,meV in (c) and
(e), respectively, where theory predicts $g\mu_{\rm B} 7\,\rm{T} =
0.81$\,meV with $g=2$ for nanotubes.
\newline\indent
We now analyze the size of hexagon (1,1), which involves 1, 2, and 3
electrons in the DQD-shell. We show that by applying a magnetic
field the 2-electron ground state can be changed from ${\rm S_B}$ to
${\rm T_{B-}}$\,, which is used to estimate the exchange energy
($J$) (energy separation between ${\rm S_B}$ and ${\rm T_{B0}}$).
Transport at the first Coulomb peak in Fig.~\ref{fig:fig3}(d) is
through different chemical potentials at low and high magnetic
field, given by \cite{hansonreview} $\mu_{\rm S_B \leftrightarrow
10} \propto +\frac{1}{2} g \mu_{\rm B} B$ at low magnetic field ($g
\mu_{\rm B} B < J$), and $\mu_{\rm T_{B-} \leftrightarrow 10}
\propto -\frac{1}{2} g \mu_{\rm B} B$ at high magnetic field ($g
\mu_{\rm B} B > J$).
Similarly, transport at the second Coulomb peak in
Fig.~\ref{fig:fig3}(d) is through $\mu_{12 \leftrightarrow \rm S_B}
\propto -\frac{1}{2} g \mu_{\rm B} B$ at low magnetic field ($g
\mu_{\rm B} B < J$) and through $\mu_{12 \leftrightarrow \rm T_{B-}}
\propto +\frac{1}{2} g \mu_{\rm B} B$ at high magnetic field ($g
\mu_{\rm B} B > J$)\cite{hansonreview}.
The same magnetic field dependence is expected for the height of
hexagon (1,1) (see Fig.~\ref{fig:fig3}(f)).
Therefore, for increasing magnetic field, hexagon (1,1) decreases in
size when ${\rm S_B}$ is ground state, and increases in size when
${\rm T_{B-}}$ is ground state, schematically shown in
Fig.~\ref{fig:fig3}(b).
The measurements in Fig.~\ref{fig:fig3}(d) and (f) are in good
agreement with the theory in Fig.~\ref{fig:fig3}(b) with the bend
(shift of ground state from singlet to triplet) occurring at $B\sim
2$ - $3$\,T, corresponding to an exchange energy of $J \sim 0.23$ -
$0.35$\,meV.
\newline\indent
The exchange energy can, for large negative detuning (center of
hexagon (1,1)), also be estimated from the tunnel coupling strength
($t$) using $J \simeq 4(t\sqrt2)^2/U_{\rm C1}$ (see supplement
material) \cite{loss,hansonreview}. We estimate $t \sim 0.32$\,meV
from the curvature of the hexagons at the anticrossings (see
supplement material) \cite{graber}. This estimate of $t$ yields a
consistent estimate of the exchange energy $J \simeq
4(t\sqrt2)^2/U_{\rm C1} \sim 0.27$\,meV.
\newline \indent
The anticrossing between (1,1) and (0,2) (red area in
fig.~\ref{fig:fig2}(b)) is analyzed in Fig.~\ref{fig:fig4}. We find
that transport is governed by elastic and inelastic cotunneling via
${\rm S_B}$ and ${\rm T_{B-}}$\,. The chemical potential for adding
an electron to ${\rm S_B}$ and ${\rm T_{B-}}$ with $E_1 + E_2 = 0$,
i.e., along the black dashed line in Fig.~\ref{fig:fig4}(a) is given
by (see supplement):
\begin{equation}\label{eq:mu_s_b}
\begin{array}{lcl}
\mu_{\rm S_{B} \leftrightarrow 01} (\varepsilon, B) & = &
-\frac{1}{2}\sqrt{(2t\sqrt2)^2 + \varepsilon^2} +
\frac{1}{2}g\mu_{\rm B} B
\end{array}
\end{equation}
\begin{equation}\label{eq:mu_t_b}
\begin{array}{lcl}
\mu_{\rm T_{B-} \leftrightarrow 01}(\varepsilon, B) & = &
-\frac{1}{2}\Big(\sqrt{(2t)^2 + (\varepsilon -
\Delta E_2)^2} - \Delta E_2 \Big) \\
 & & - \frac{1}{2}g\mu_{\rm B} B
\end{array}
\end{equation}
We plot Eq.~(\ref{eq:mu_s_b}) and (\ref{eq:mu_t_b}) in
Fig.~\ref{fig:fig4}(b) with $B=0$\,T (solid green and blue lines),
and with $B=6$\,T (dashed green and blue lines). We see that ${\rm
S_B}$ is ground state for $B=0$\,T, and that the two chemical
potentials cross at elevated magnetic field, indicated with red
arrow.
\newline \indent
At low magnetic field one broad peak in conductance versus detuning
between (1,1) and (0,2) is seen (Fig.~\ref{fig:fig4}(a), white arrow
marked A). This conductance peak is due to elastic cotunneling via
${\rm S_B}$, schematically shown in Fig.~\ref{fig:fig4}(c) (mark A).
Since elastic cotunneling via ${\rm S_B}$ involves both S(11) and
S(02), which have equal weight at $\varepsilon = 0$, the elastic
cotunneling peak is centered around $\varepsilon = 0$. At high
magnetic field the elastic cotunneling via ${\rm S_B}$ becomes
suppressed because the ground state at $\varepsilon = 0$ changes
from ${\rm S_B}$ to ${\rm T_{B-}}$\,. Fig.~\ref{fig:fig4}(d) shows a
surface plot of $I_{\rm sd}$ versus $\varepsilon$ and $B$ along the
black dashed line in Fig.~\ref{fig:fig4}(a). The white vertical line
marked A is the expected position of the elastic cotunneling.
\newline \indent
At high magnetic field we observe two narrow peaks, marked B and C
in Fig.~\ref{fig:fig4}(a). These two narrow peaks are due to the
onset of inelastic cotunneling via ${\rm S_B}$ and ${\rm T_{B-}}$\,,
schematically shown in Fig.~\ref{fig:fig4}(c) mark B and C. Onset of
inelastic cotunneling via ${\rm S_B}$ and ${\rm T_{B-}}$ occurs when
the energy separation between their chemical potentials becomes
equal to the applied bias:
\begin{equation}\label{eq:condition}
e V_{\rm sd} = \pm \big( \mu_{\rm S_{B} \leftrightarrow 01}
(\varepsilon, B) - \mu_{\rm T_{B-} \leftrightarrow 01}(\varepsilon,
B) \big)
\end{equation}
We have from these two conditions calculated the onset of inelastic
cotunneling in $(\varepsilon, B)$-space and plotted it as white
lines marked B and C in Fig.~\ref{fig:fig4}(d). Note that no fitting
parameters are used in Fig.~\ref{fig:fig4}(d), the parameters used,
$t=0.32$\,meV, $\Delta E_2 =1.5$\,meV were found in the analysis
above.
\newline \indent
\section{methods}
\subsection{Fabrication and measurement setup}
The devices are made on a highly doped silicon substrate with a top
layer of silicon dioxide. The CNTs are grown by chemical vapor
deposition from islands of catalyst material and subsequently
contacted by 50\,nm Titanium source and drain electrodes. Next,
three narrow top-gate electrodes are fabricated between the source
and drain electrodes, consisting of aluminum oxide and titanium
\cite{jorgensen}. A schematic figure of the device together with the
measurement setup is shown in Fig.~\ref{fig:fig1}(a). Source-drain
voltage ($V_{\rm sd}$) is applied to the source electrode and the
drain electrode is connected through a current-to-voltage amplifier
to ground. The three top-gate electrodes are named G1, CG (center
gate), and G2 starting from the source electrode. For the device
reported on in this Letter we saw that G1 had a much lower
gate-coupling than G2 and CG (see Fig.~\ref{fig:fig1}(b) and
Fig.~\ref{fig:fig2}(a)) which we attribute to the G1-electrode being
damaged somewhere, weakening its gate-coupling. The gate coupling of
G1 to dot 1 is $\alpha_{G1}=2.9$\,meV/V, and gate coupling of G2 to
dot 2 is $\alpha_{G2}=400$\,meV/V. The center gate is kept at
$V_{\rm CG}=0V$ for the measurements shown in this Letter. All data
presented in this Letter are measured in a sorption pumped $^3$He
cryostat at 350\,mK.
\section{Acknowledgements}
We wish to acknowledge the support of the EU-STREP ULTRA-1D program
and the EU FP6 CANAPE project.
\section{Competing financial interests}
The authors declare no competing financial interests.
\bibliographystyle{nat}
\end{document}